\documentclass[a4paper]{article}

\usepackage{INTERSPEECH_v2}
\usepackage{graphicx}
\usepackage{amsmath}
\usepackage{amssymb}
\usepackage{textcomp}
\usepackage{hyperref}
\usepackage[ruled,vlined]{algorithm2e}
\usepackage{color}
\usepackage{cite}

\ninept

\DeclareGraphicsExtensions{.pdf,.png,.jpg}

\title{A modulation property of time-frequency derivatives of filtered phase and \\
its application to aperiodicity and $f_{\rm o}$ estimation\thanks{%
The main body of this article is accepted for publication in Interspeech2017.
This article has supplemental materials for details which were dropped due to limited space.}}
\name{Hideki Kawahara$^1$, Ken-Ichi Sakakibara$^2$, Masanori Morise$^3$, Hideki Banno$^4$, Tomoki Toda$^5$}
\address{
 $^1$Wakayama University, Japan\\
 $^2$Health Science University of Hokkaido, Japan\\
 $^3$University of Yamanashi, Japan\\
 $^4$Meijo University, Japan\\
 $^5$Graduate School of Information Science, Nagoya University, Japan}
 \email{kawahara@sys.wakayama-u.ac.jp, kis@hoku-iryo-u.ac.jp, mmorise@yamanashi.ac.jp, banno@meijo-u.ac.jp, tomoki@icts.nagoya-u.ac.jp}

\begin{document}

\maketitle
\begin{abstract}
We introduce a simple and linear SNR (strictly speaking, periodic to random power ratio)
 estimator (0~dB to 80~dB without additional calibration/linearization) for providing reliable descriptions of
aperiodicity in speech corpus.
The main idea of this method is to estimate the background random noise level without directly extracting the background noise.
The proposed method is applicable to a wide variety of time windowing functions with very low sidelobe levels.
The estimate combines the frequency derivative and the time-frequency derivative of the mapping from filter center frequency to the output instantaneous frequency.
This procedure can replace the periodicity detection and aperiodicity estimation subsystems of
recently introduced open source vocoder, YANG vocoder.
Source code of MATLAB implementation of this method will also be open sourced.
\end{abstract}
\noindent\textbf{Index Terms}: speech analysis, fundamental frequency, aperiodicity, instantaneous frequency, group delay

\section{Introduction}
Aperiodic components in speech sound play important roles in speech communication (normal, expressive, extreme and so on) and singing expression\cite{fujisaki1997prosody,Sakakibara2001VocalFA,Gobl2003spcom,fujimura2009jlpv}.
They contribute to the synthesized speech quality and intelligibility\cite{fujimura1968ieeetr,makhoul1978mixed,griffin1988ieeetr,yoshimura2001mixed}.
In spite of their importance, reliable estimation of aperiodic components has been a challenging topic\cite{yegnanarayana1998ieeetr,Kawahara1999,Jackson2001PitchscaledEO,Kawahara2001,deshumukh2005ieetr}. 
The major difficulty is the fact that such aperiodic components in speech are generally much weaker than
the periodic components.
We propose {\em to use deviations of the prominent periodic components} to estimate the level of the aperiodic components\cite{Kawahara1999}, {\em instead of directly extracting and measuring them}.

This article is organized as follows.
First, we briefly review aperiodicity estimation methods and their underlying principles.
Second, we intuitively introduce the underlying idea of the proposed method. 
Third, we introduce and discuss on each component of the proposed method.
They are, stabilization of instantaneous frequency estimation,
modulation behavior of mapping from the filter center frequency to the output instantaneous frequency,
and effects of the maximum sidelobe level and the decaying behavior of sidelobes.
Then, we introduce  and test several implementation models and their performance.
Finally, we discuss the issues of applying the proposed method to
speech analysis / synthesis systems.

The main contribution of this paper is a simple SNR estimator based on the above-mentioned idea. 
The estimator uses a six-term cosine series 
function, 
which was proposed to implement the antialiased Fujisaki-Ljungqvist model\cite{kawahara2017ISAAFL}. 
This 
function provides a linear and robust SNR estimation 
ranging from 0~dB to 80~dB SNR without additional linearization.

\section{Background}

Voiced sounds are almost periodic but are not perfectly periodic\cite{fujimura1968ieeetr}.
Amplitude and timing of each glottal cycle are not exactly the same each other.
It also comprises random noise, owing to turbulence around constrictions of vocal tract and glottis.
Among these factors, amplitude and timing deviations can be estimated using spectrum analysis and
$f_{\rm o}$\footnote{We use ``$f_{\rm o}$'' instead of ``F0'' based on the recommendation\cite{titze2015jasaforum}.
}
 (fundamental frequency) analysis, respectively\cite{deshumukh2005ieetr,Malyska2008,kawahara2016using}.
In this paper, we focus on the remaining difficult issue, the random noise. 
We also focus on its spectral shape, because our primary goal is to use them for designing excitation signals for speech synthesis and modification.

One typical approach for estimating random noise component is to extract random components using harmonic analysis or using residuals of inverse filtering\cite{fujisaki1987icassp,stylianou1995high,yegnanarayana1998ieeetr,dalessandro1998ieee,Kawahara2001,deshumukh2005ieetr,Alku2011}.
Prediction and subtraction of waveforms one pitch-period apart,
also try to measure the residual levels\cite{cheveigne2002jasa,deshumukh2005ieetr,Kawahara2010IS}.
These approaches are generally sensitive to the $f_{\rm o}$ estimation error and the model parameter errors, because of the large level difference between prominent periodic component and the random noise component.
In other words, removing the effects of the strong periodic component is the key issue to be solved.

We introduced an algorithm to eliminate the prominent periodic component, without estimating the frequency of the periodic component\cite{kawahara2016using}.
The algorithm alleviates the issue mentioned above based on a self-tuning process, which does not require prior frequency information.
This merit of the algorithm is obtained at the expense of temporal resolution.
In this paper, we propose an alternative approach, which does not estimate the random components directly and
does not need to sacrifice the temporal resolution.
It starts from revisiting our legacy-STRAIGHT\cite{kawahara1999spcom} and the fixed-point-based $f_{\rm o}$ extractor\cite{Kawahara1999}.


\section{Phase-deviation-based SNR estimation}
The instantaneous frequency of a sinusoidal component, which is separated by using a band-pass filter, is modulated by other components contained in the same filter output.
The relative background noise level to the prominent periodic component can be estimated by measuring the magnitude of this modulation.
This idea was implied in our fixed-point-based $f_{\rm o}$ extractor\cite{Kawahara1999} but was not investigated further.
We revisit this idea in this paper based on 
detailed analyses of the filtered signal phase behavior and the windowing function.

We focus on phase derivatives, instead of using the phase directly.
Derivatives of the phase of a filtered signal are easier to interpret than phase itself.
The time derivative of the phase provides the instantaneous frequency and
the frequency derivative (with negative sign) provides the group delay.
The power-spectrum-weighted average of each representation
provides centroid of the frequency and the temporal position, respectively\cite{Boashash1992,Boashash1992a,cohen95}.
These are better understood by deriving Flanagan's instantaneous frequency equation\cite{flanagan66jasa} and 
similarly derived group delay equation\cite{kawahara2000icslp}.

When a band-pass filter comprises one prominent sinusoidal component and a low-level interfering sinusoid,
the instantaneous frequency of the filter output varies at the rate of their frequency difference.
The magnitude of the variation is proportional to the ratio of their amplitude.
This ratio depends on the relative location of the filter center frequency and the components.
In other words, the magnitude of the frequency derivative of the mapping from the filter center frequency to the output instantaneous frequency provides the estimate of the ratio of their amplitude.
For a sinusoidally varying signal, the signal and its time derivative are orthogonal to each other and the magnitude is represented as the RMS (root mean square) value with proper scaling.
A random noise is a collection of independent identically distributed frequency components, the RMS value of the frequency derivative and the time-frequency derivative of the above-mentioned mapping provides the ratio of the prominent sinusoidal component and the background random component.
This is an intuitive introduction to the proposed method.

For the proposed method to work properly, the following issues to be solved.
\begin{description}
\item[Removal of singularities] Cancellation of neighboring harmonic components yields singularities of instantaneous frequency.
Effects of background noise are significantly magnified in the vicinity of these singularities and are harmful for reliable estimation.
\item[Rejection of leakage from outside] Voiced signal consists of harmonic components.
The target band-pass filter has to isolate one harmonic component from the others.
The other harmonic components have to be sufficiently attenuated. 
\item[Scaling of derivatives] For minimizing the effect of the leakage and statistical fluctuations, the signal and the time derivative of the signal have to be properly scaled.
\item[Design and calibration] The estimator has performance governing system parameters.
The estimator design needs selecting and tuning these parameters, followed by calibration using known test signals.
\end{description}
The following sections address and solve each problem.

\subsection{Removal of singularities}
To remove singularities, revisit the derivation of the instantaneous frequency.
Assume that the impulse response of a band-pass filter is complex valued and the filtered output $x(t) = |x(t)| e^{j\theta(t)}$ is also complex valued.
The imaginary part of the logarithmic conversion of $x(t)$ is the phase function $\theta(t)$.
The time derivative of it yields the Flanagan's phase equation\cite{flanagan66jasa}.
By definition, it is the instantaneous (angular) frequency $\omega_i(t)$:
\begin{align}
\omega_i(t) 
& = \frac{\Re[x(t)]\dfrac{d\Im[x(t)]}{dt}-\Im[x(t)]\dfrac{d\Re[x(t)]}{dt}}{|x(t)|^2} ,\label{ew:flanaganIF}
\end{align}
where the denominator is zero yields singularity.

Let us define the filter impulse response $h(t, \omega)$ by using a symmetric non-negative windowing function $w(t)$ for its envelope and using a complex exponential $e^{j\omega t}$ for the carrier signal.
Therefore, the filter output is  a function of $t$ and the carrier frequency $\omega$ and is represented using $x(t, \omega)$.
The instantaneous frequency of this filter output $\omega_i(t, \omega)$ is represented as follows:
{\small
\begin{align}
\!\!\!\omega_i(t, \omega) & = \!
\frac{\Re[x(t,\omega)]\Im[x_d(t,\omega)]\!\!-\!\!
 \Im[x(t,\omega)]\Re[x_d(t,\omega)]}{|x(t,\omega)|^2} , \label{eq:filteredFNG}
\end{align}}
where $x_d(t,\omega)$ represents the time derivative of $x(t,\omega)$.
Because the filter output $x(t, \omega)$ is the convolution of the input signal $s(t)$ and the impulse response $h(t, \omega)$, the time derivative of $x(t, \omega)$ is represented as follows:
\begin{align}
x_d(t, \omega) & = j\omega h(t, \omega) \ast s(t) - h_d(t, \omega) \ast s(t) , \label{eq:twoFilters}
\end{align}
where ``$\ast $'' represents convolution and $h_d(t, \omega)$ represents the time derivative of $h(t, \omega)$.
Equations~\ref{eq:filteredFNG} and \ref{eq:twoFilters} provide a procedure for calculating instantaneous frequency without introducing phase unwrapping and numerical differentiation.

Equation~\ref{eq:filteredFNG} indicates that making denominator always positive removes singularities.
Specifically, using  $|x(t, \omega)|^2$ for the weight and using a non-negative smoothing kernel $w_F(\omega)$ to calculate the following weighted average $\omega_{is}(t, \omega)$ removes singularities\cite{Boashash1992,Boashash1992a}.
\begin{align}
\omega_{is}(t, \omega) & = \frac{\int_{-\infty}^{\infty} w_F(\nu) |x(t, \omega-\nu)|^2 \omega_i(t, \omega-\nu) d\nu}{\int_{-\infty}^{\infty} w_F(\nu) |x(t, \omega-\nu)|^2 d\nu} ,
\end{align}
where the kernel $w_F(\omega)$ is bounded inside $(-\omega_w, \omega_w)$.

\subsection{Rejection of leakage from outside}
Because speech signals are time varying, the observation window has to be bounded in time.
This finite bounding results in infinite spread in the frequency domain.
The prolate spheroidal wave function provides the best time-frequency concentration among time bounded functions\cite{slepian1961prolate,slepian1978prolate}.
The Kaiser window\cite{kaiser1980use} is an engineering approximation of this best function.

The frequency domain representation of bounded functions has the main lobe and the sidelobes.
The normalized frequency bandwidth characterizes the main lobe.
The maximum level and the decay rate characterize the sidelobes.
These indices can be designed and have to be tested in terms of their effects on the SNR estimation performance.
Cosine series windowing functions\cite{harris1978ieee,nattall1981ieee,kawahara2017ISAAFL} are useful for testing trade-offs between these design indices.

We proposed a six-term cosine window\cite{kawahara2017ISAAFL} for antialiased Fujisaki-Ljungqvist model and it turned out that the proposed six-term window also is the best function for  SNR estimation.
It has strong attenuation (-114.24~dB) and steep decay rate (54~dB/oct) of sidelobes.
The form of the window $w(t)$ and the coefficients are as follows:\footnote{%
The coefficients are rounded to ten digits to the right of the decimal point. 
The best coefficients were selected from rounded numbers.
Derivation details are given in our paper\cite{kawahara2017ISAAFL}.}
\begin{align}
w(t) & = \sum_{k=0}^5 h_k \cos\left( \frac{k \pi t}{t_w} \right) \\
\!\{h_k\}_{k=0}^5 & =  \left\{0.2624710164, 0.4265335164, 0.2250165621,  \right. \nonumber \\
& \!\!\! \left. 0.0726831633, 0.0125124215, 0.0007833203\right\} , \label{eq:sixTerm}
\end{align}
where $t_w$ represents the half length of the windowing function and for $|t| \ge t_w$, the value of $w(t)$ is zero.

\subsection{Scaling of derivatives}
The variance of the frequency derivative of the mapping from the filter center frequency to the output instantaneous frequency, $V_{\Delta \omega}= V\!\left[\frac{d \omega_{is}(t, \omega)}{d \omega}\right]$, is approximately proportional to the following value:
\begin{align}
C_{\Delta \omega} & = \frac{\int_{-\omega_w}^{\omega_w} |W(\omega) \omega|^2 d\omega }{
\int_{-\omega_w}^{\omega_w} |W(\omega)|^2 d\omega},
\end{align}
where $W(\omega)$ represents the frequency response of the response envelope $w(t)$.
The variance of the time-frequency derivative $V_{\Delta \omega t}=V\!\left[\frac{d^2 \omega_{is}(t, \omega)}{d \omega dt}\right]$ is also approximately proportional to:
\begin{align}
C_{\Delta \omega t} & =\frac{ \int_{-\omega_w}^{\omega_w} |W(\omega) \omega^2|^2 d\omega}{
\int_{-\omega_w}^{\omega_w} |W(\omega) |^2 d\omega} .
\end{align}

Scaling each derivative to make $V_{\Delta \omega}=V_{\Delta \omega t}$ minimizes the variance of the mixed index $\eta(t, \omega)$ defined as follows:
\begin{align}
\eta(t, \omega) & = \left[\left(\frac{d \omega_{is}(t, \omega)}{d \omega}\right)^2 + 
 C_{eq} \left(\frac{d^2 \omega_{is}(t, \omega)}{d \omega dt}\right)^2\right]^{\frac{1}{2}} ,
\end{align}
where $C_{eq} \propto C_{\Delta \omega} / C_{\Delta \omega t}$ represents the scaling coefficient.

\subsection{Design and calibration}\footnote{MATLAB scripts and functions used here are online-available. \\
http://www.sys.wakayama-u.ac.jp/\%7ekawahara/IS2017AP/}
We conducted a set of simulations to set design parameters, the windowing function, the window length, the smoothing kernel and the size, and the scaling coefficients.
The first test is for trade-offs between the maximum sidelobe level and the decay rate of sidelobes.
Table~\ref{winfunc} shows the tested windowing functions.
The items named ``DPSS-4'' and ``DPSS-4.5'' represent the windows based on the prolate spheroidal wave function, 
which is implemented as a MATLAB function \texttt{dpss}.
The table lists the characterizing indices of each windowing function.
The length (nominal length) of each window was designed so as to locate the first zero point at $f_{\rm o}$.
\begin{table}[tbp]
\caption{Tested windowing functions. Windows 1 to 5 are cosine serie\cite{harris1978ieee,nattall1981ieee,kawahara2017ISAAFL}. 
Windows 7 and 8 are the best energy bounded functions\cite{slepian1961prolate,slepian1978prolate}.
The window 6 is their approximation\cite{kaiser1980use}.}
\vspace{-0.3cm}
{\small
\begin{center}
\begin{tabular}{l|ccl}
\hline
 & maximum & asymptotic & \multicolumn{1}{c}{nominal} \\
 & sidelobe level & decay rate &  \multicolumn{1}{c}{length} \\
\multicolumn{1}{c|}{Type} &  (dB) &  (dB/oct) &  \multicolumn{1}{c}{(re. $1/f_{\rm o}$)} \\
\hline\hline
1 Hanning &  \hspace{0.08cm} -31.47 & 18 &   \hspace{0.4 cm} 2 \\
2 Blackman &  \hspace{0.08cm} -58.11 & 18 &  \hspace{0.4 cm} 3\\
3 Nuttall-11 &  \hspace{0.08cm} -82.60 & 30 &  \hspace{0.4 cm} 4\\
4 Nuttall-15 & \hspace{0.08cm} -98.17 &  \hspace{0.08cm} 6 &  \hspace{0.4 cm} 4\\
5 six-term & -114.24 & 54 &  \hspace{0.4 cm} 6\\
6 Kaiser &  \hspace{0.08cm} -98.32 &  \hspace{0.08cm} 6 &  \hspace{0.4 cm} 4.273 \\
7 DPSS-4 & \hspace{0.08cm}  -95.47 &  \hspace{0.08cm} 6 &  \hspace{0.4 cm} 4.078 \\
8 DPSS-4.5 & -108.57 &  \hspace{0.08cm} 6 &  \hspace{0.4 cm} 4.567  \\
\hline
\end{tabular}
\end{center}}
\label{winfunc}
\vspace{-0.3cm}
\end{table}%

The test signal was a 100~Hz pulse train plus Gaussian white noise with given SNR.
The sampling frequency was set 44,100~Hz.
The length of the FFT buffer was 32,768 samples.
We tested at all possible relative window center locations in each pitch cycle.
We also tested at all harmonic frequencies up to the Nyquist frequency and yielded 96,579 estimates for each test condition.
Each window was scaled 1.5 times of its nominal length and
the width of the frequency smoother $w_F(\omega)$\footnote{We used a raised cosine for $w_F(\omega)$ in this implementation. 
Other smoothing functions are being tested.} 
was set a value from 0.25 to 0.35, based on preliminary tests.
Each estimator was calibrated using the median of the distribution of the mixed index $\eta(t, \omega)$.

\begin{figure}[tbp]
\begin{center}
\includegraphics[width=0.99\hsize]{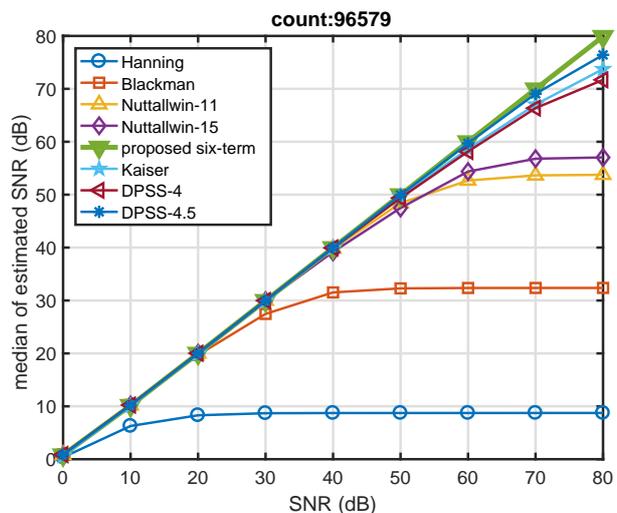}
\vspace{-0.3cm}
\caption{SNR estimation using cosine series windows.
The horizontal axis shows the given SNR and the vertical axis shows the
median of the estimated SNR. }
\label{apEst}
\end{center}
\vspace{-0.3cm}
\end{figure}
Figure~\ref{apEst} shows the relation between the given SNR and the median of estimated SNR using the windows in Table~\ref{winfunc}.
The results suggest that the most contributing factor on the estimation accuracy is the maximum sidelobe level.
The proposed six-term cosine series shows the best linearity from 0~dB to 80~dB SNR range.
Other windows show saturation in the estimated SNR.

{
\begin{figure}[tbp]
\begin{center}
\includegraphics[width=0.99\hsize]{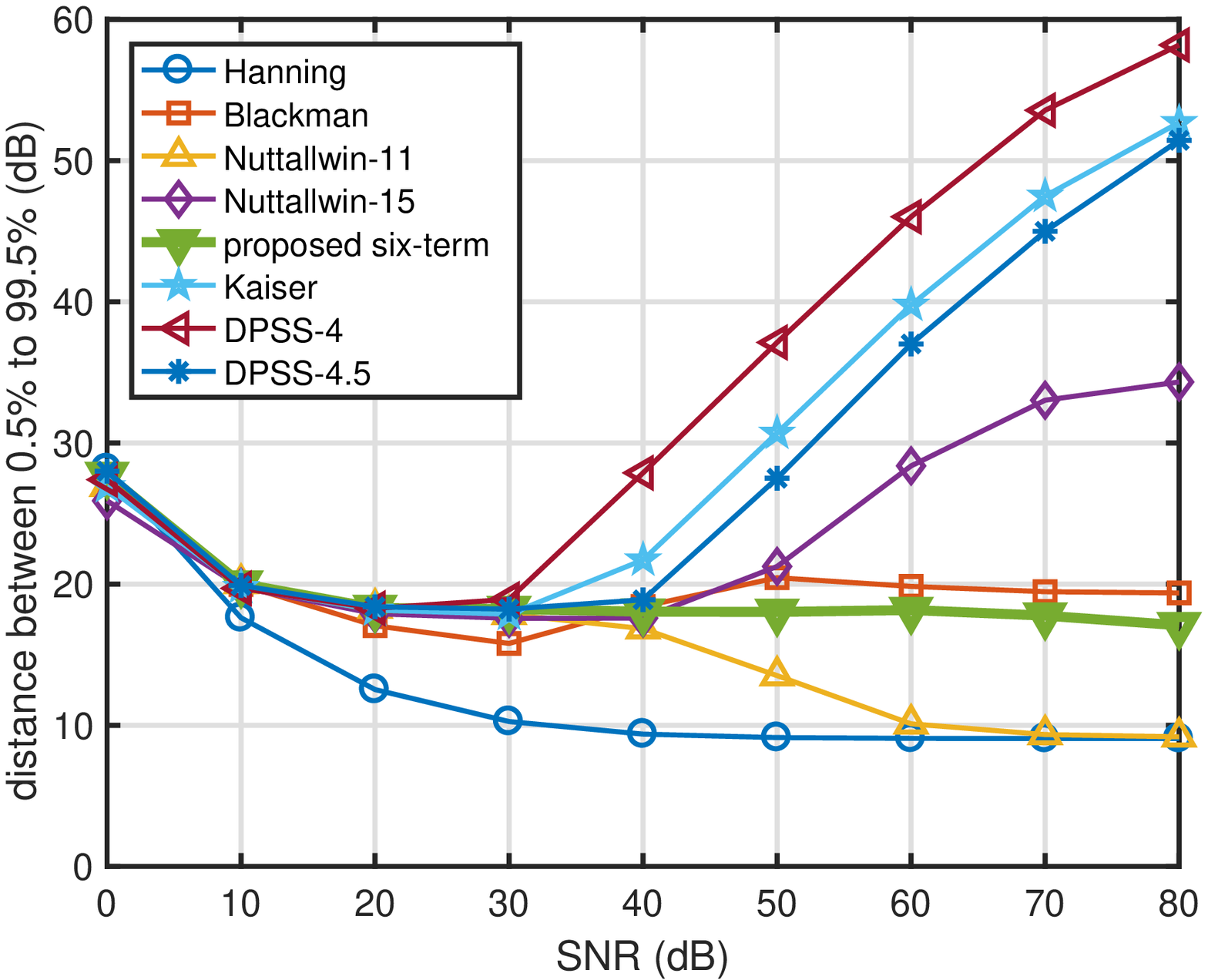}
\vspace{-0.3cm}
\caption{Distribution of SNR estimation.
The vertical axis shows the distance between 0.5\% and 99.5\% locations of the cumulative distribution
of the estimated SNR.}
\label{spread}
\end{center}
\begin{center}
\includegraphics[width=0.99\hsize]{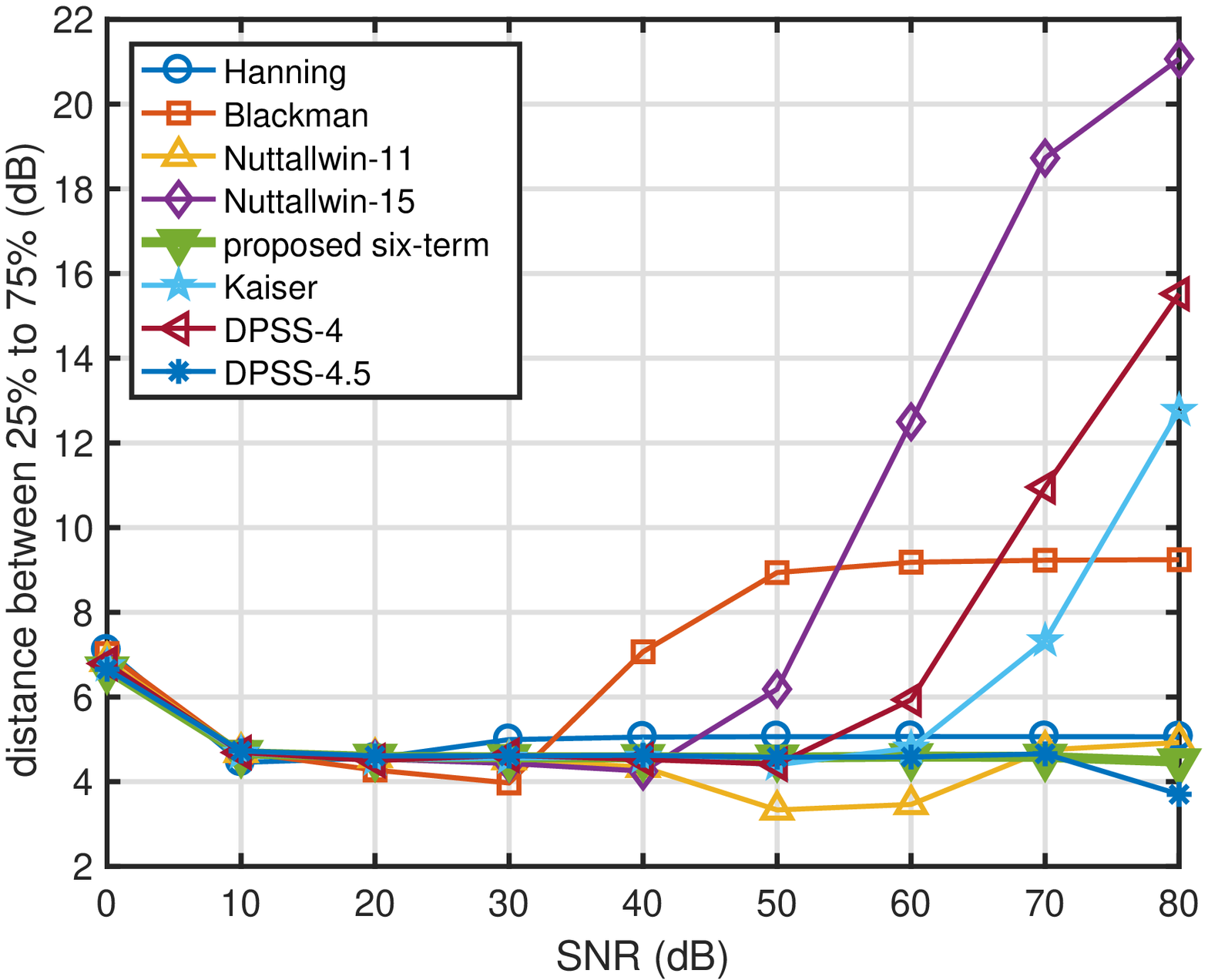}
\vspace{-0.3cm}
\caption{Distribution of SNR estimation.
The vertical axis shows the distance between 25\% and 75\% locations of the cumulative distribution
of the estimated SNR.
}
\label{spreadW}
\end{center}
\vspace{-0.5cm}
\end{figure}
}

Figure~\ref{spread} and  \ref{spreadW} illustrate the behavior of outliers of the estimates.
Figure~\ref{spread} shows the distance between 0.5\% and 99.5\% points of each cumulative distribution.
It illustrates that the windows with 6~dB/oct sidelobe decay rate have extreme outliers.
The smaller width of distribution for Hanning and Nuttallwin-11 is caused by the saturation of the estimates and does not indicate that the windows perform better.

Figures~\ref{spreadW} shows that Blackman window suffers from distorted distribution for SNR higher than 40~dB.
In addition, the estimates by Blackman window saturate from 30~dB SNR.
These results indicate that the proposed six-term cosine series window provides the best estimates.
The second best is Nuttallwin-11, because saturation from 50~dB is practically acceptable and it has shorter window length (nominal length: 4) than the six-term window (nominal length: 6).

\section{Application to real speech analysis}
The proposed procedure, when it is implemented using FFT, works properly for known constant $f_{\rm o}$ signals.
However, $f_{\rm o}$ of real speech varies almost all the time.
To make apparent $f_{\rm o}$ of such signal constant,
we apply an adaptive time axis warping using the fundamental frequency trajectory $f_{\rm o}(t)$ of the signal and the target constant frequency $f_{\rm tgt}$.
The ratio $f_{\rm o}(t)/f_{\rm tgt}$ provides the necessary stretching factor of the time axis.
This adaptive time warping also removes side-band components, which are the results of frequency modulation of harmonic components\cite{kobayashi1997if,Kawahara1999,Malyska2008}.
Algorithm~\ref{apForReal} shows the pseudo code of this procedure,
where $K$ represents the number of frames and $N(k)$ represents the number of harmonic components at the $k$-th frame.
The coefficient $C_0$ represents the calibration constant, which depends on the used window function.
\begin{algorithm}[tbp] 
  \SetAlgoLined
  \KwData{Speech signal $s(t)$ and $f_{\rm o}(t)$}
  \KwResult{Time-frequency map of SNR, $SNR(t, \omega)$}
  Adaptive time warp: $s(t) \longrightarrow u(\tau)$ using $\frac{f_{\rm o}(t)}{f_{\rm tgt}}$ \;
  Convert the original frame time $t_k$ to the frame time on the warped time $\tau_k$ \;
  \For{$k \leftarrow 1$ \KwTo $K$ }{
    Spectrum analysis: $u(\tau)$ using $w(t)$ and $w_d(t)$ at $\tau_k$\;
    \For{$n \leftarrow 1 \KwTo N(k) $}{
    $\nu_n \leftarrow 2 \pi nf_{\rm tgt}$ \;
    $SNR(\tau_k,  \nu_n) \leftarrow C_0 \eta( \tau_k, \nu_n)$ \;
    $\omega_n \leftarrow nf_{\rm o}(t_k)$ \;
    $SNR(t_k, \omega_n) \leftarrow SNR(\tau_k, \nu_n)$ \;
    }
    }
    Interpolate $\{SNR(t_k, \omega_n)\} \rightarrow SNR(t, \omega)$ \;
  \caption{SNR estimation of real speech}
  \label{apForReal}
\end{algorithm}

\subsection{Example and issues for excitation source design}
\begin{figure}[tbp]
\begin{center}
\includegraphics[width=0.99\hsize]{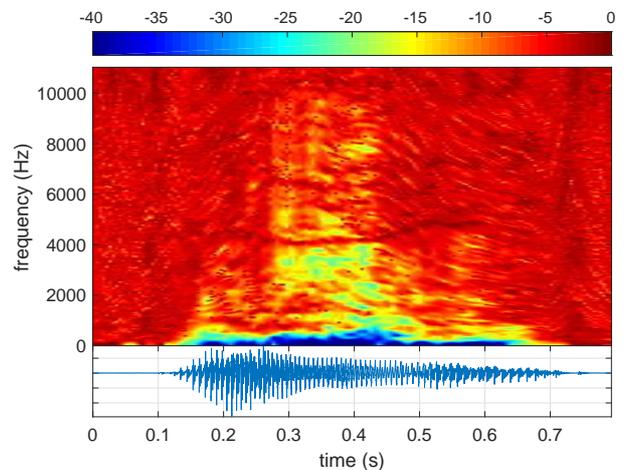}
\vspace{-0.3cm}
\caption{Estimated time-frequency SNR map of a Japanese vowel sequence /aiueo/ spoken by a male speaker.
This map used the proposed six-term function.
The bottom panel shows the waveform.
The middle panel shows the SNR map. The value of $-20\log_{10}(SNR(t, \omega))$ is displayed using the pseudo-color mapping, which is shown as the color bar on the top panel.
}
\label{vaiueoAPgram}
\end{center}
\vspace{-0.3cm}
\end{figure}
Figure~\ref{vaiueoAPgram} shows an example SNR map of a Japanese vowel sequence /aiueo/, spoken by a male speaker.
The signal was sampled at 22,050~Hz with 16~bit resolution.
The SNR in dB value is clipped at 40~dB.

The estimated SNR $SNR(t, \omega)$ is sufficient to describe the signal characteristics.
It is useful for generating annotation of physical attribute to speech corpus, for example.
However, it is not enough to design an excitation source signal for synthesizing speech.
There are four factors to be considered; non-speech background noise, the discrepancy between $f_{\rm o}(t)$ movement and movement of resonance frequencies of the vocal tract, and the auditory filter shape.

The background noise dominates when/where the power of the speech signal is weak.
Figure~\ref{vaiueoAPgram} has a dark horizontal curved trace around 4,500~Hz from 0.2~s to 0.6~s.
The trace roughly corresponds to the deep spectral dip, owing to pyriform fossa\cite{takemoto2010acoustic}.
This dip of speech power may contribute to this dark trace.
The background noise also introduces bias in the estimated SNR.
This effect has to be compensated before determining the noise level in the excitation source signal.

The discrepancy between $f_{\rm o}(t)$ movement and movement of resonance frequencies of the vocal tract makes the estimated SNR increase at resonance frequencies.
It is because the resonance frequencies on the warped time axis move and deviate from precise repetition.
These resonances have to be equalized to alleviate this effect before estimating SNR.

The initial stage of the human auditory system is approximately a constant-Q filter bank.
Each filter bandwidth is proportional to its center frequency.
The SNR map has to be 
smoothed according to this frequency resolution
for designing the excitation source signal.
This also reduces the variance of estimates.

In addition to these, non-linear interactions of the vocal tract and the vocal fold vibration, with its temporal fluctuation\cite{titze2008Jasa} may introduce apparent aperiodicity around resonant frequencies.
Detailed analysis and tuning of the proposed method to the target application, specifically speech synthesis, are issues which need further research.

\section{Discussion}
The proposed mixed index $\eta(t, \omega)$ can also be implemented using wavelet analysis, similar to YANG vocoder\cite{kawahara2016using}.
The SNR estimator based on wavelet analysis can replace the periodicity detector in the first stage of YANG vocoder and can provide the probability map of $f_{\rm o}$ location.
This way, the proposed method is applicable to the initial stage of $f_{\rm o}$ estimation.
The FFT-based implementation is also applicable to the refinement stage of $f_{\rm o}$ estimation of YANG vocoder.

We did not discuss the temporal distribution of the random component within one pitch period, but it is an important issue\cite{Jackson2001PitchscaledEO,Malyska2008}.
It has to be properly estimated
because the temporal location of a noise burst has a significant impact on the perceived noise level\cite{Skoglund2000ieeetrans}.
A weighted average of group delay with minimum phase compensation\cite{kawahara2000icslp} can be applicable to this issue by revisiting derivation of the group delay equation and windowing function design.
This is the next research target of aperiodicity analysis.

\section{Conclusion}
We propose a simple and linear SNR (periodic to noise component ratio) estimator, which is operational spanning from 0~dB to 80~dB SNR range, without using additional linearization.
We introduced a six-term cosine series window, which has very low maximum sodelobe level (-114.24~dB) and steep decay rate (54~dB/oct), for implementing the estimator.
These will serve as a dependable infrastructure for speech analysis.
We are planning to use this estimator for updating our new open source vocoder (YANG vocoder).

\section{Acknowledgements}
This work was supported by JSPS KAKENHI Grant Numbers~JP15H03207, JP15H02726 and JP16K12464.

{\small
\bibliographystyle{IEEEtran}

\bibliography{IS2017KawaharaAP}

\begin{thebibliography}{10}
\providecommand{\url}[1]{#1}
\csname url@samestyle\endcsname
\providecommand{\newblock}{\relax}
\providecommand{\bibinfo}[2]{#2}
\providecommand{\BIBentrySTDinterwordspacing}{\spaceskip=0pt\relax}
\providecommand{\BIBentryALTinterwordstretchfactor}{4}
\providecommand{\BIBentryALTinterwordspacing}{\spaceskip=\fontdimen2\font plus
\BIBentryALTinterwordstretchfactor\fontdimen3\font minus
  \fontdimen4\font\relax}
\providecommand{\BIBforeignlanguage}[2]{{%
\expandafter\ifx\csname l@#1\endcsname\relax
\typeout{** WARNING: IEEEtran.bst: No hyphenation pattern has been}%
\typeout{** loaded for the language `#1'. Using the pattern for}%
\typeout{** the default language instead.}%
\else
\language=\csname l@#1\endcsname
\fi
#2}}
\providecommand{\BIBdecl}{\relax}
\BIBdecl

\bibitem{fujisaki1997prosody}
H.~Fujisaki, ``{Prosody, Models, and Spontaneous Speech},'' in \emph{Computing
  Prosody}.\hskip 1em plus 0.5em minus 0.4em\relax New York, NY: Springer US,
  1997, pp. 27--42.

\bibitem{Sakakibara2001VocalFA}
K.-I. Sakakibara, H.~Imagawa, T.~Kouishi, K.~Kondo, E.~Z. Murano, M.~Kumada,
  and S.~Niimi, ``Vocal fold and false vocal fold vibrations in throat singing
  and synthesis of kh{\"{o}{\"{o}mei}},'' in \emph{ICMC}, 2001.

\bibitem{Gobl2003spcom}
C.~Gobl and A.~{N{\'{i}} Chasaide}, ``{The role of voice quality in
  communicating emotion, mood and attitude},'' \emph{Speech Communication},
  vol.~40, no. 1-2, pp. 189--212, 2003.

\bibitem{fujimura2009jlpv}
O.~Fujimura, K.~Honda, H.~Kawahara, Y.~Konparu, M.~Morise, and J.~C. Williams,
  ``{Noh voice quality.}'' \emph{Logopedics, phoniatrics, vocology}, vol.~34,
  no.~4, pp. 157--170, 2009.

\bibitem{fujimura1968ieeetr}
O.~Fujimura, ``An approximation to voice aperiodicity,'' \emph{IEEE
  Transactions on Audio and Electroacoustics}, vol.~16, no.~1, pp. 68--72, Mar
  1968.

\bibitem{makhoul1978mixed}
J.~Makhoul, R.~Viswanathan, R.~Schwartz, and A.~Huggins, ``A mixed-source model
  for speech compression and synthesis,'' \emph{The Journal of the Acoustical
  Society of America}, vol.~64, no.~6, pp. 1577--1581, 1978.

\bibitem{griffin1988ieeetr}
D.~W. Griffin and J.~S. Lim, ``Multiband excitation vocoder,'' \emph{IEEE
  Transactions on Acoustics, Speech, and Signal Processing}, vol.~36, no.~8,
  pp. 1223--1235, Aug 1988.

\bibitem{yoshimura2001mixed}
T.~Yoshimura, K.~Tokuda, T.~Masuko, T.~Kobayashi, and T.~Kitamura, ``Mixed
  excitation for hmm-based speech synthesis.'' in \emph{INTERSPEECH}, 2001, pp.
  2263--2266.

\bibitem{yegnanarayana1998ieeetr}
B.~Yegnanarayana, C.~{d'Alessandro}, and V.~Darsinos, ``{An iterative algorithm
  for decomposition of speech signals into periodic and aperiodic
  components},'' \emph{IEEE Transactions on Speech and Audio Processing},
  vol.~6, no.~1, pp. 1--11, jan 1998.

\bibitem{Kawahara1999}
H.~Kawahara, H.~Katayose, A.~de~Cheveign{\'{e}}, and R.~D. Patterson, ``{Fixed
  point analysis of frequency to instantaneous frequency mapping for accurate
  estimation of F0 and periodicity.}'' in \emph{Proc. Eurospeech 99}, Budapest,
  Hungary, 1999, pp. 2781--2784.

\bibitem{Jackson2001PitchscaledEO}
P.~J.~B. Jackson and C.~H. Shadle, ``Pitch-scaled estimation of simultaneous
  voiced and turbulence-noise components in speech,'' \emph{IEEE Trans. Speech
  and Audio Processing}, vol.~9, pp. 713--726, 2001.

\bibitem{Kawahara2001}
H.~Kawahara, J.~Estill, and O.~Fujimura, ``{Aperiodicity extraction and control
  using mixed mode excitation and group delay manipulation for a high quality
  speech analysis, modification and synthesis system STRAIGHT},'' in
  \emph{Proceedings of MAVEBA}, Firentze Italy, 2001, pp. 59--64.

\bibitem{deshumukh2005ieetr}
O.~Deshmukh, C.~Y. Espy-Wilson, A.~Salomon, and J.~Singh, ``Use of temporal
  information: Detection of periodicity, aperiodicity, and pitch in speech,''
  \emph{IEEE Transactions on Speech and Audio Processing}, vol.~13, no.~5, pp.
  776--786, Sept 2005.

\bibitem{kawahara2017ISAAFL}
H.~Kawahara, K.-I. Sakakibara, H.~Banno, M.~Morise, T.~Toda, and T.~Irino, ``{A
  new cosine series antialiasing function and its application to aliasing-free
  glottal source models for speech and singing synthesis},'' in
  \emph{Interspeech 2017}, Stockholm, 2017, [Accepted].

\bibitem{titze2015jasaforum}
I.~R. Titze, R.~J. Baken, K.~W. Bozeman, S.~Granqvist, N.~Henrich, C.~T.
  Herbst, D.~M. Howard, E.~J. Hunter, D.~Kaelin, R.~D. Kent, J.~Kreiman,
  M.~Kob, A.~L{\"{o}}fqvist, S.~McCoy, D.~G. Miller, H.~No{\'{e}}, R.~C.
  Scherer, J.~R. Smith, B.~H. Story, J.~G. {\v{S}}vec, S.~Ternstr{\"{o}}m, and
  J.~Wolfe, ``{Toward a consensus on symbolic notation of harmonics,
  resonances, and formants in vocalization},'' \emph{The Journal of the
  Acoustical Society of America}, vol. 137, no.~5, pp. 3005--3007, 2015.

\bibitem{Malyska2008}
N.~Malyska and T.~F. Quatieri, ``{Spectral representations of nonmodal
  phonation},'' \emph{IEEE Transactions on Audio, Speech and Language
  Processing}, vol.~16, no.~1, pp. 34--46, 2008.

\bibitem{kawahara2016using}
\BIBentryALTinterwordspacing
H.~Kawahara, Y.~Agiomyrgiannakis, and H.~Zen, ``{Using instantaneous frequency
  and aperiodicity detection to estimate F0 for high-quality speech
  synthesis},'' \emph{arXiv preprint arXiv:1605.07809}, 2016. [Online].
  Available: \url{http://arxiv.org/abs/1605.07809}
\BIBentrySTDinterwordspacing

\bibitem{fujisaki1987icassp}
H.~Fujisaki and M.~Ljungqvist, ``{Estimation of voice source and vocal tract
  parameters based on ARMA analysis and a model for the glottal source
  waveform},'' in \emph{ICASSP 1987}, 1987, pp. 637--640.

\bibitem{stylianou1995high}
Y.~Stylianou, J.~Laroche, and E.~Moulines, ``High-quality speech modification
  based on a harmonic+ noise model,'' in \emph{Fourth European Conference on
  Speech Communication and Technology}, 1995, pp. 451--454.

\bibitem{dalessandro1998ieee}
C.~{d'Alessandro}, V.~Darsinos, and B.~Yegnanarayana, ``{Effectiveness of a
  periodic and aperiodic decomposition method for analysis of voice sources},''
  \emph{IEEE Transactions on Speech and Audio Processing}, vol.~6, no.~1, pp.
  12--23, jan 1998.

\bibitem{Alku2011}
P.~Alku, ``{Glottal inverse filtering analysis of human voice production - A
  review of estimation and parameterization methods of the glottal excitation
  and their applications},'' \emph{Sadhana -- Academy Proceedings in
  Engineering Sciences}, vol.~36, no. October, pp. 623--650, 2011.

\bibitem{cheveigne2002jasa}
A.~de~Chevengn{\'{e}} and H.~Kawahara, ``{YIN, a fundamental frequency
  estimator for speech and music},'' \emph{The Journal of the Acoustical
  Society of America}, vol. 111, no.~4, pp. 1917--1930, 2002.

\bibitem{Kawahara2010IS}
H.~Kawahara, H.~Itagaki, Y.~Wada, M.~Morise, R.~Nisimura, and T.~Irino,
  ``{Analysis and synthesis of singing with hoarse vocal expressions},'' in
  \emph{Proceedings of the 2010 Annual Conference of the Australian Acoustical
  Society}, vol.~5, 2010.

\bibitem{kawahara1999spcom}
H.~Kawahara, I.~Masuda-Katsuse, and A.~de~Cheveigne, ``{Restructuring speech
  representations using a pitch-adaptive time-frequency smoothing and an
  instantaneous-frequency-based F0 extraction},'' \emph{Speech Communication},
  vol.~27, no. 3-4, pp. 187--207, 1999.

\bibitem{Boashash1992}
B.~Boashash, ``{Estimating and interpreting the instantaneous frequency of a
  signal. I. Fundamentals},'' \emph{Proceedings of the IEEE}, vol.~80, no.~4,
  pp. 520--538, 1992.

\bibitem{Boashash1992a}
------, ``{Estimating and Interpreting the Instantaneous Frequency of a Signal
  - Part 2: Algorithms and Applications},'' \emph{Proceedings of the IEEE},
  vol.~80, no.~4, pp. 540--568, 1992.

\bibitem{cohen95}
L.~Cohen, \emph{{Time-frequency analysis}}.\hskip 1em plus 0.5em minus
  0.4em\relax Englewood Cliffs, NJ: Prentice Hall, 1995.

\bibitem{flanagan66jasa}
J.~L. Flanagan and R.~M. Golden, ``{Phase Vocoder},'' \emph{Bell System
  Technical Journal}, vol.~45, no.~9, pp. 1493--1509, nov 1966.

\bibitem{kawahara2000icslp}
H.~Kawahara, Y.~Atake, and P.~Zolfaghari, ``{Accurate vocal event detection
  method based on a fixed-point analysis of mapping from time to weighted
  average group delay},'' in \emph{ICSLP 2000}, 2000, pp. 664--667.

\bibitem{slepian1961prolate}
D.~Slepian and H.~O. Pollak, ``{Prolate spheroidal wave functions, Fourier
  analysis and uncertainty-I},'' \emph{Bell System Technical Journal}, vol.~40,
  no.~1, pp. 43--63, 1961.

\bibitem{slepian1978prolate}
D.~Slepian, ``{Prolate spheroidal wave functions, Fourier analysis, and
  uncertainty-V: The discrete case},'' \emph{Bell System Technical Journal},
  vol.~57, no.~5, pp. 1371--1430, 1978.

\bibitem{kaiser1980use}
J.~Kaiser and R.~W. Schafer, ``{On the use of the $I_0$-sinh window for
  spectrum analysis},'' \emph{Acoustics, Speech and Signal Processing, IEEE
  Transactions on}, vol.~28, no.~1, pp. 105--107, 1980.

\bibitem{harris1978ieee}
F.~J. Harris, ``{On the use of windows for harmonic analysis with the discrete
  Fourier transform},'' \emph{Proceedings of the IEEE}, vol.~66, no.~1, pp.
  51--83, 1978.

\bibitem{nattall1981ieee}
A.~H. Nuttall, ``{Some windows with very good sidelobe behavior},'' \emph{IEEE
  Trans. Audio Speech and Signal Processing}, vol.~29, no.~1, pp. 84--91, 1981.

\bibitem{kobayashi1997if}
T.~Abe, T.~Kobayashi, and S.~Imai, ``The {IF} spectrogram: {A} new spectral
  representation,'' in \emph{Proceedings of International Symposium on
  Simulation, Visualization and Auralization for Acoustics Research and
  Education}, 1997, pp. 423--430.

\bibitem{takemoto2010acoustic}
H.~Takemoto, P.~Mokhtari, and T.~Kitamura, ``{Acoustic analysis of the vocal
  tract during vowel production by finite-difference time-domain method},''
  \emph{The Journal of the Acoustical Society of America}, vol. 128, no.~6, pp.
  3724--3738, 2010.

\bibitem{titze2008Jasa}
I.~R. Titze, ``{Nonlinear source--filter coupling in phonation: Theory},''
  \emph{The Journal of the Acoustical Society of America}, vol. 123, no.~5, pp.
  2733--2749, may 2008.

\bibitem{Skoglund2000ieeetrans}
J.~Skoglund and W.~B. Kleijn, ``{On time-frequency masking in voiced speech},''
  \emph{Speech and Audio Processing, IEEE Transactions on}, vol.~8, no.~4, pp.
  361--369, jul 2000.

\end{thebibliography}
}


\appendix
\section{Windowing functions}
\begin{figure}[htbp]
\begin{center}
\hfill\includegraphics[width=0.89\hsize]{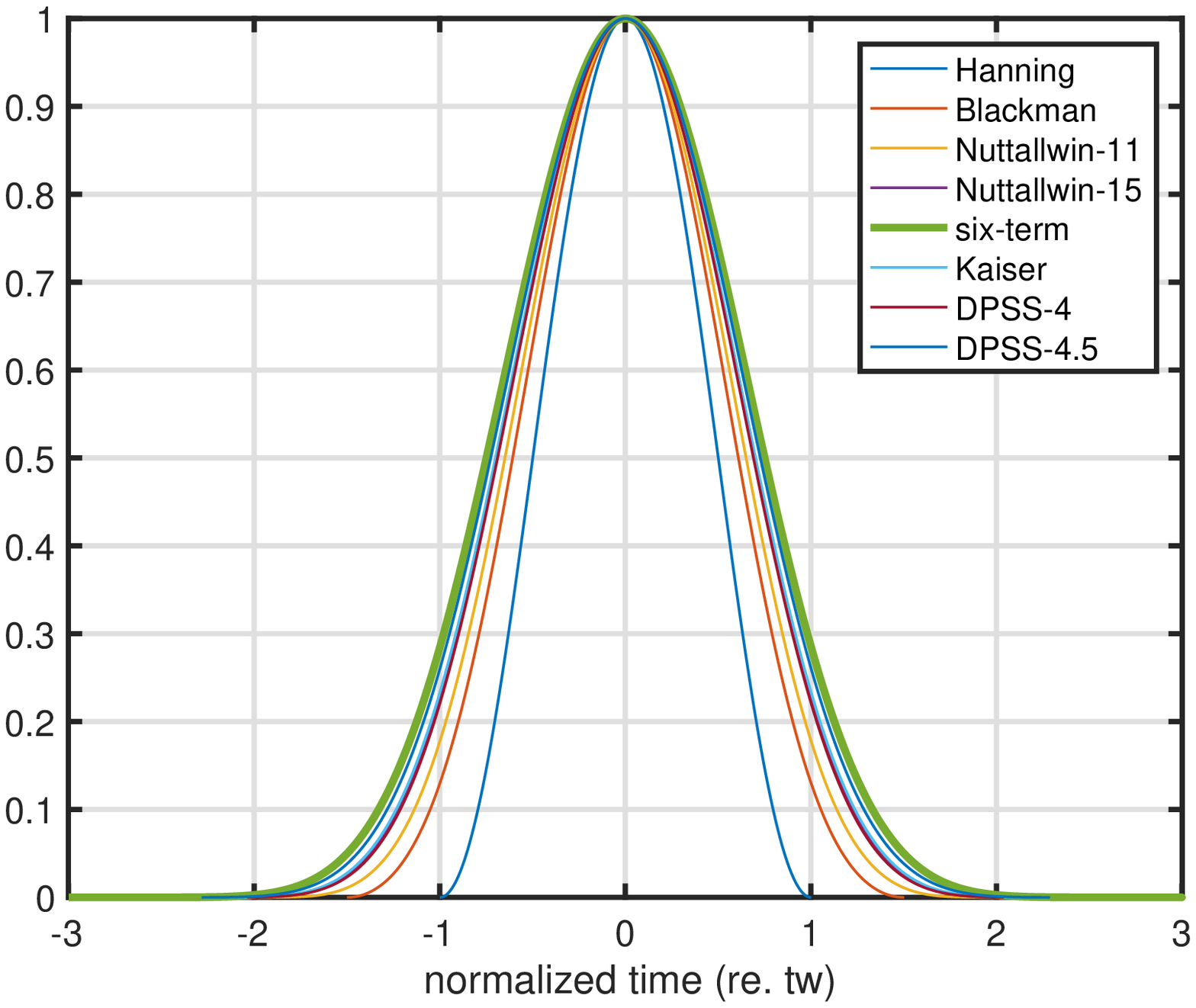}\\
\includegraphics[width=0.99\hsize]{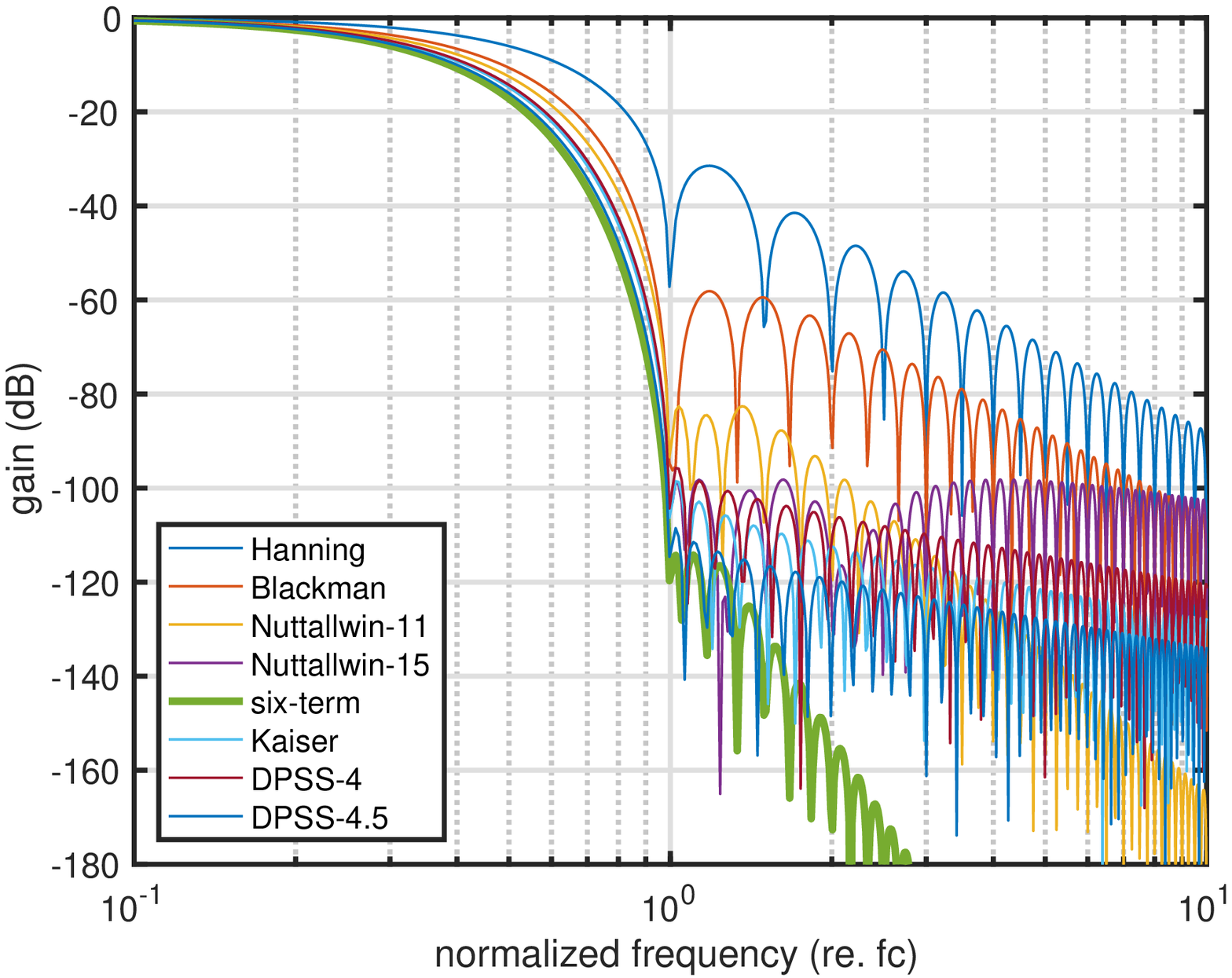}
\caption{Windowing function shapes and gain. }
\label{winGainNominal}
\end{center}
\end{figure}
Figure~\ref{winGainNominal} shows the temporal shape and
the frequency gain characteristics of used windowing functions.
The upper plot shows the shape and the bottom plot shows
the gain.
The nominal length of each windowing function shown in Table~\ref{winfunc} is
designed to make the first zero to locate at one on the normalized frequency.

However, the nominal length is not relevant for representing windowing functions.
It is relevant to represent their duration and bandwidth in terms of 2nd order moment,
$\sigma_t$ and $\sigma_f$ respectively.
\begin{align}
\sigma_t^2 & = \int_{-\infty}^{\infty} |w(t)|^2 t^2 dt \\
\sigma_f^2 & = \int_{-\infty}^{\infty} |W(f)|^2 f^2 df ,
\end{align}
where the temporal representation $|w(t)|^2$ and the frequency representation
$|W(f)|^2$ are normalized for their integration to yield one.

\begin{table}[tbp]
\caption{Duration and band-width of windowing functions}
\begin{center}
\begin{tabular}{l|cccc}
\hline
\multicolumn{1}{c|}{Type} & $\sigma_t$ & $\sigma_t/\sigma_{tR}$ & $\sigma_f$ & $\sigma_f/\sigma_{fR}$ \\
\hline\hline
1 Hanning & 0.2829 & 0.9800 & 0.2861 & 0.4959 \\
2 Blackman & 0.3564 & 1.2346 & 0.2216 & 0.3841 \\
3 Nuttall-11 & 0.3828 & 1.3261 & 0.2057 & 0.3565 \\
4 Nuttall-15 & 0.4106 & 1.4223 & 0.1915 & 0.3320 \\
5 six-term & 0.4444 & 1.5394 & 0.1766 & 0.3061 \\
6 Kaiser & 0.4142 & 1.4349 & 0.1898 & 0.3289 \\
7 DPSS-4 & 0.4072 & 1.4105 & 0.1931 & 0.3348 \\
8 DPSS-4.5 & 0.4299 & 1.4892 & 0.1827 & 0.3167 \\ \hline
\end{tabular}
\end{center}
\label{durAndBw}
\end{table}%
Table~\ref{durAndBw} shows the duration and the bandwidth of
each windowing function,
where $\sigma_{tR}$ represents the duration of a rectangular window of the unit length
and $\sigma_{fR}$ represents the bandwidth of a rectangular low-pass filter of
unit cutoff frequency.

\section{Distribution of estimated SNR}
This section provides detailed simulation results
which were used to generate Fig.~\ref{apEst}, Fig.~\ref{spread} and Fig.~\ref{spreadW}.
The SNR of the test signal varied from 0~dB to 80~dB in 10~dB steps.

\begin{figure}[tbp]
\begin{center}
\includegraphics[width=0.99\hsize]{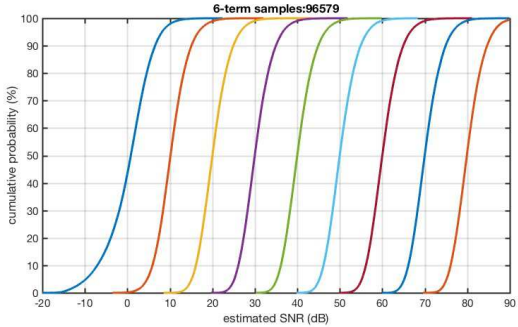}
\caption{Cumulative distribution of the true SNR level and the estimated SNR. 
The windowing function is the proposed six-term cosine series.}
\label{ifmov03960sixterm}
\end{center}
\end{figure}
Figure~\ref{ifmov03960sixterm} shows the simulation results using
the proposed six-term cosine series window.
The horizontal axis represents the estimated SNR and the
vertical axis represents the probability of the
estimated SNR $SNR_{\mathrm E}$ yields lower value than the horizontal axis value $SNR_{\mathrm Th}$.
In short, the vertical axis represents $P_r(SNR_{\mathrm E} \le SNR_{\mathrm Th})$,
where $P_r(Q)$ represents the probability of logical predicate $Q$ to be true.

\begin{figure}[tbp]
\begin{center}
\includegraphics[width=0.99\hsize]{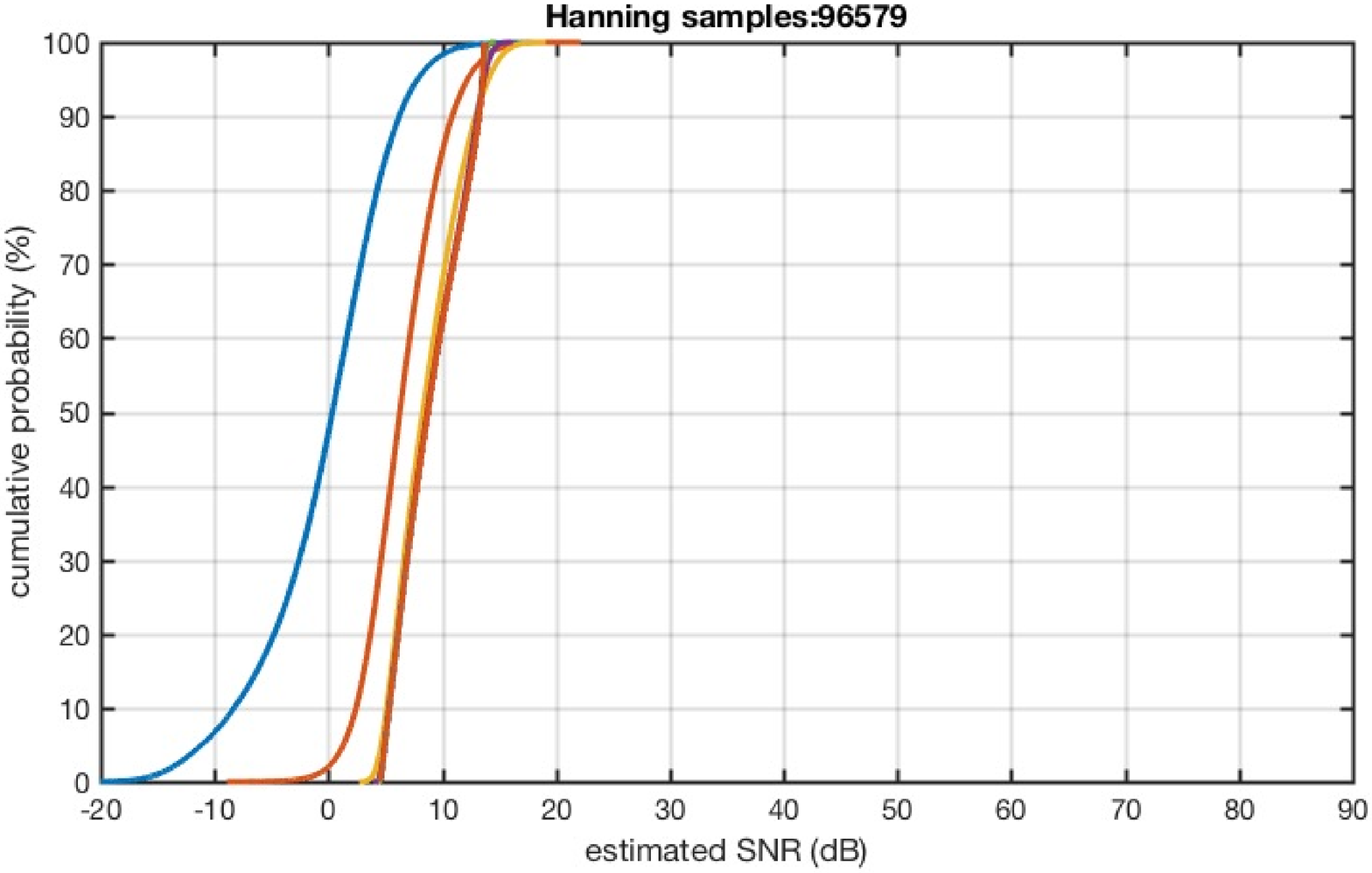}
\includegraphics[width=0.99\hsize]{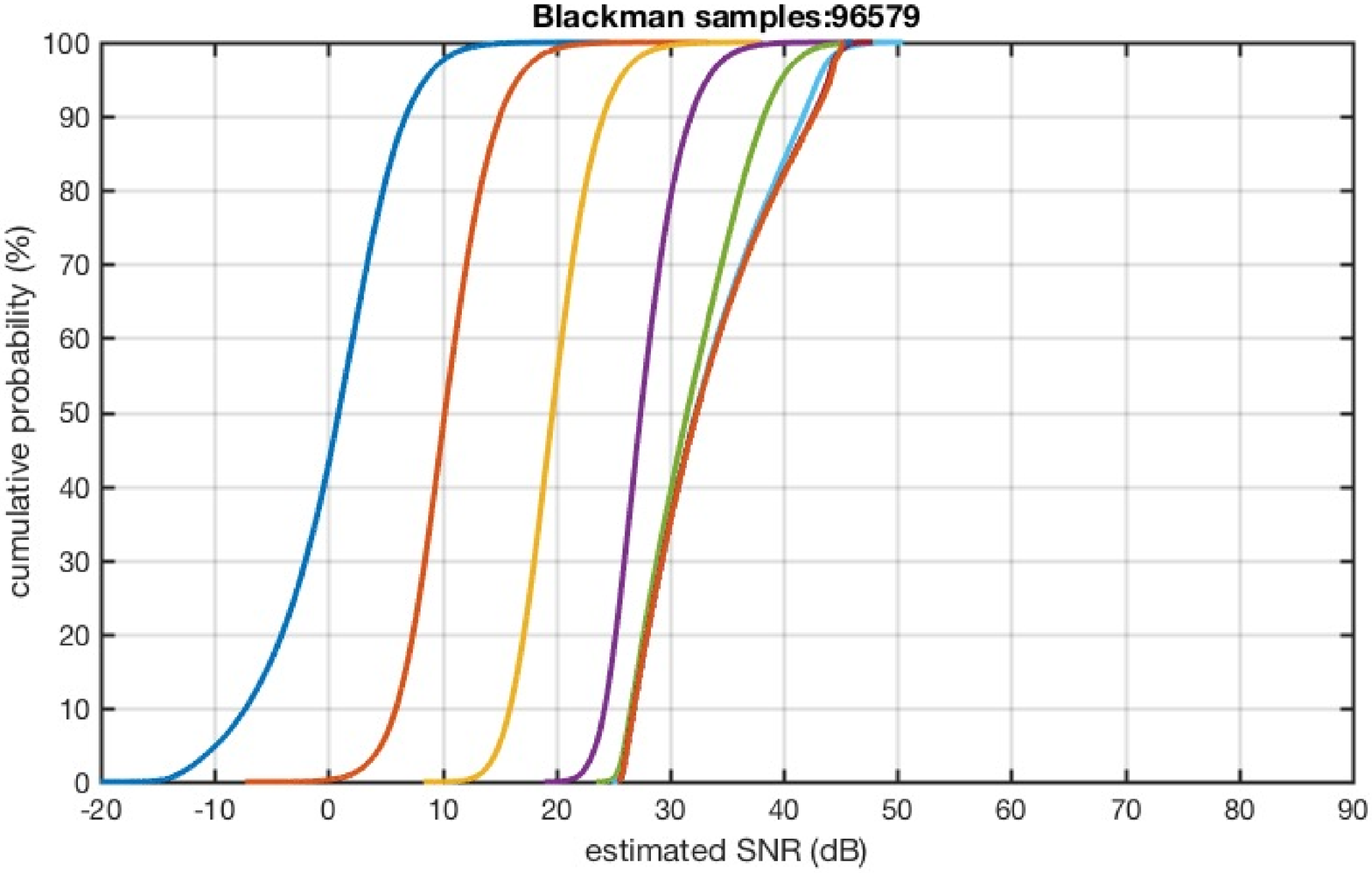}
\caption{Cumulative distribution of the true SNR level and the estimated SNR. 
Upper plot shows results using Hanning window and the lower plot shows
results using Blackman window.}
\label{ifmov03960Hanning}
\end{center}
\end{figure}
Figure~\ref{ifmov03960Hanning} shows results using Hanning and Blackman windowing functions.
Strong leakage from sidelobes made distribution for higher SNR signals saturate.

\begin{figure}[tbp]
\begin{center}
\includegraphics[width=0.99\hsize]{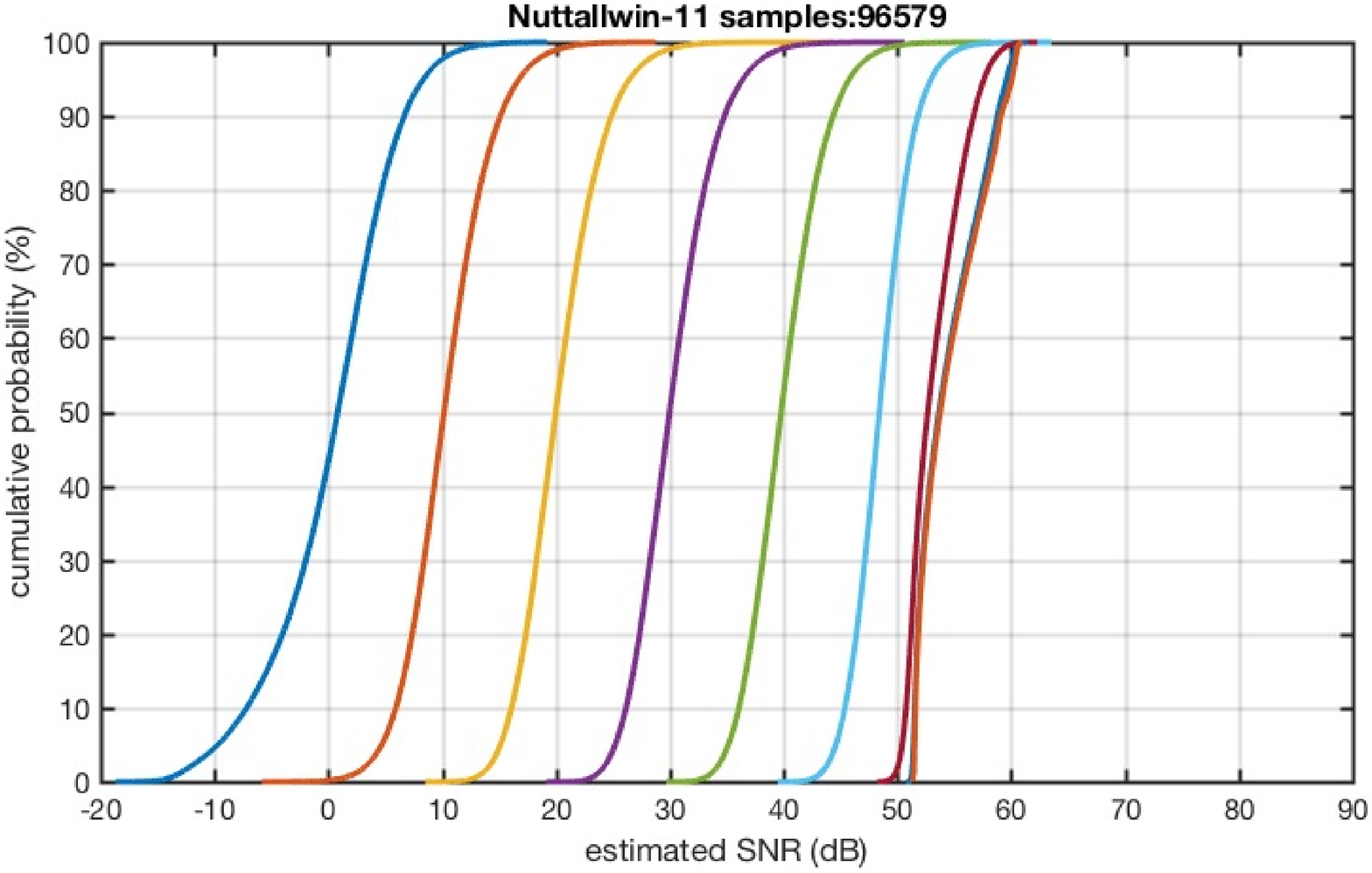}
\includegraphics[width=0.99\hsize]{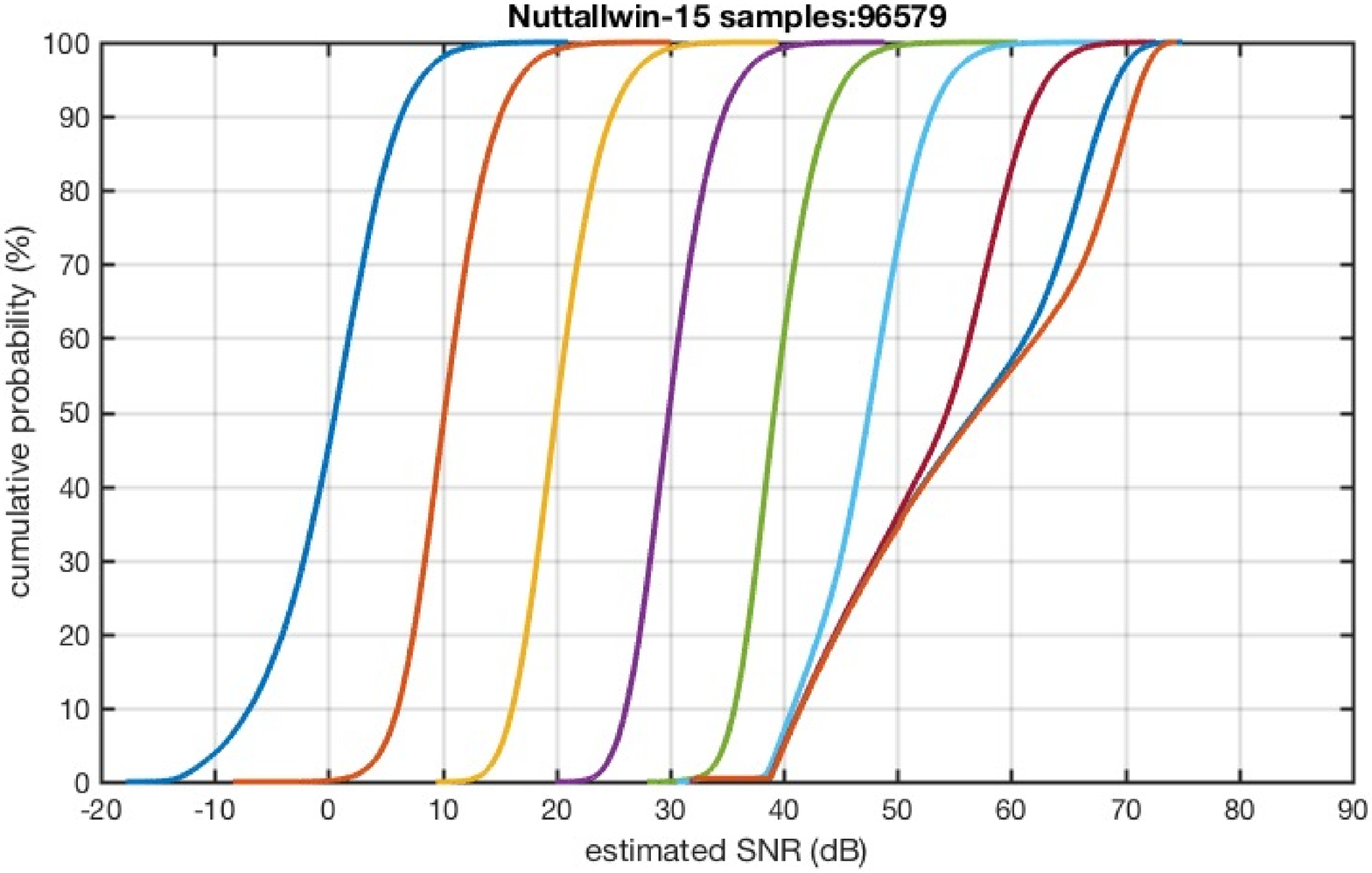}
\caption{Cumulative distribution of the true SNR level and the estimated SNR.
Upper plot shows the results using the 11-th item of the Table~II of Nuttall's windows
and the lower plot shows that for the 15-th item.
Note that MATLAB uses the 15-th item for its {\texttt nuttallwin}  function.}
\label{ifmov03960Nuttall}
\end{center}
\end{figure}
Figure~\ref{ifmov03960Nuttall} shows reults using windows with more
attenuated sidelobe levels.
The highest sidelobe level of the 15-th item is lower than the 11-th item.
For the 11-th item, saturation of the distribution occurs for 60~dB or higher SNR.
For the 15-th item, strong distortion of the distribution occurs for 60~dB or higher SNR.

\begin{figure}[tbp]
\begin{center}
\includegraphics[width=0.99\hsize]{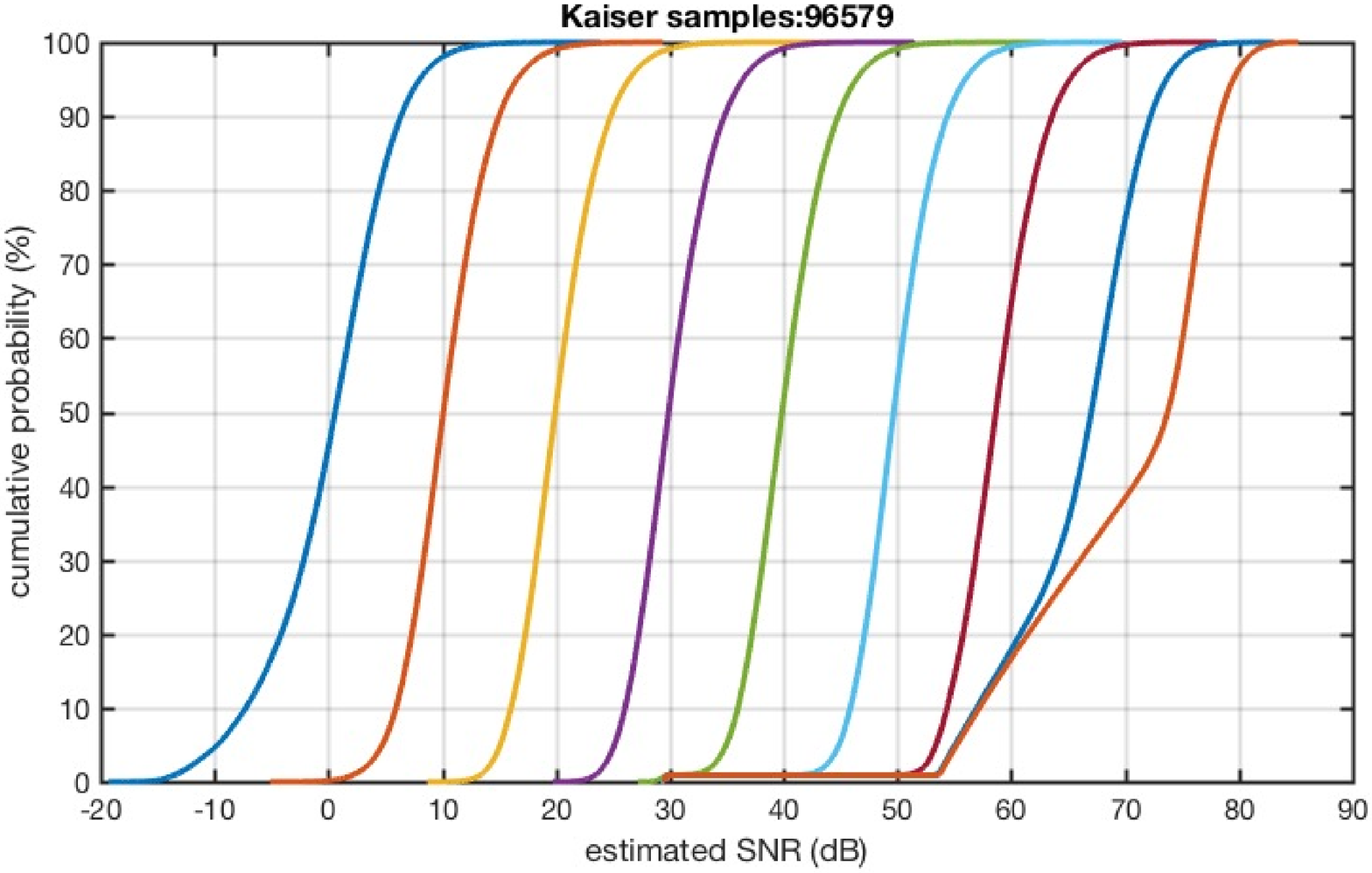}
\includegraphics[width=0.99\hsize]{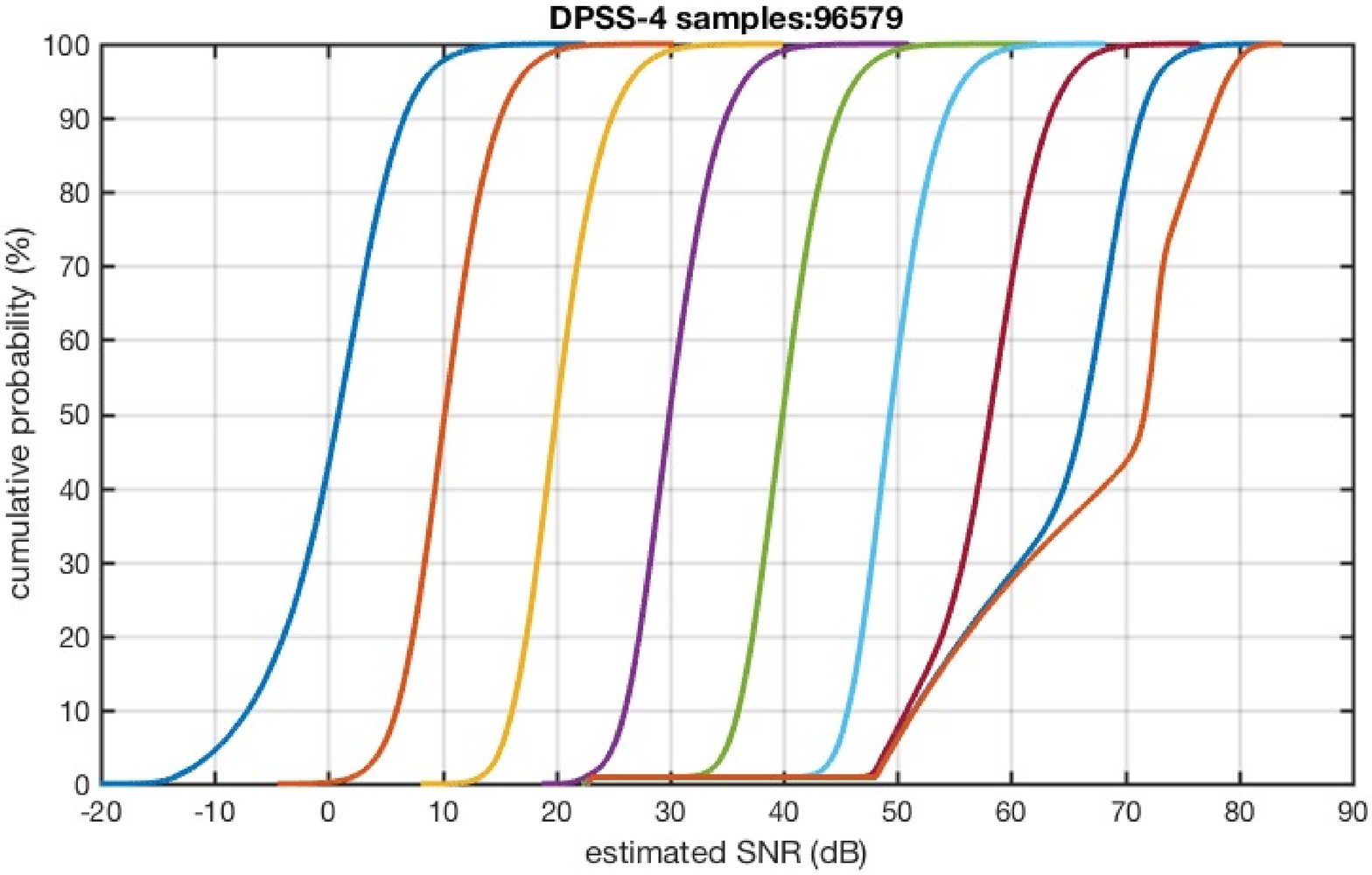}
\includegraphics[width=0.99\hsize]{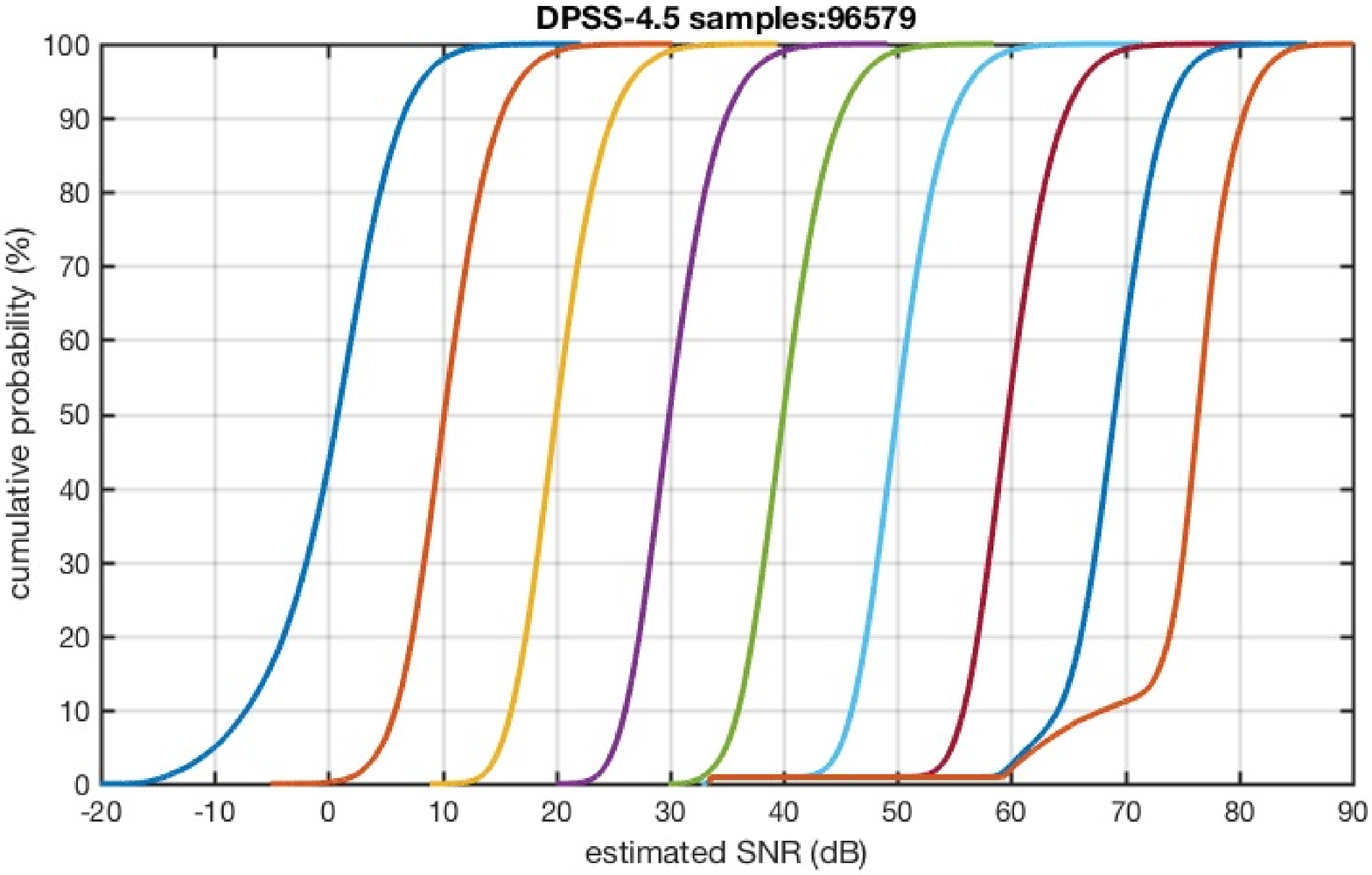}
\caption{Cumulative distribution of the true SNR level and the estimated SNR. 
The top plot shows the results using Kaiser window
and the bottom two plot shows results using prolate spheroidal wave function.
The lower plot has the similar highest sidelobe level to Kaiser window.}
\label{ifmov03960kaiser}
\end{center}
\end{figure}
Figure~\ref{ifmov03960kaiser} shows the results using
theoretically the best localized function, prolate spheroidal wave function 
and its approximation, Kaiser window.

\end{document}